# Isocurvature and Adiabatic Fluctuations of Axion in Chaotic Inflation Models and Large Scale Structure

M. Kawasaki

*Institute for Cosmic Ray Research, The University of Tokyo, Tanashi 188, Japan*
*: kawasaki@icrhp3.icrr.u-tokyo.ac.jp*

Naoshi Sugiyama

*Department of Physics & Research Center for the Early Universe, School of Science,*
*The University of Tokyo, Tokyo 113, Japan : sugiyama@yayoi.phys.s.u-tokyo.ac.jp*

T. Yanagida

*Department of Physics, School of Science, The University of Tokyo, Tokyo 113, Japan*

(December 19, 1995)

In the chaotic inflation models, quantum fluctuations for axion fields lead to the overproduction of domain walls and too large isocurvature fluctuations which is inconsistent with the observations of cosmic microwave background anisotropies. These problems are solved by assuming a very flat potential for the Peccei-Quinn scalar. As the simplest possibility, we consider a model where the Peccei-Quinn scalar is an inflaton itself and show that the isocurvature fluctuations can be comparable with the adiabatic ones. We investigate cosmological implications in the case that both adiabatic and isocurvature fluctuations exist and find that the amplitude of the matter spectrum becomes smaller than that for the pure adiabatic case. This leads to relatively high bias parameter ($b \simeq 2$) which is favoured by the current observations.

## I. INTRODUCTION

The axion [1–4] is the Nambu-Goldstone boson associated with the breaking of Peccei-Quinn symmetry which was invented as a natural solution to the strong CP problem in QCD [5]. The Peccei-Quinn symmetry breaking scale $F_a$ is stringently constrained by laboratory experiments, astrophysics and cosmology; the allowed range of $F_a$ is between $10^{10}$ GeV and $10^{12}$ GeV [6] in the standard cosmology. Axion is also cosmologically attractive since it can be cold dark matter if $F_a$ takes higher values in the allowed region.

Inflationary universe [7,8] is invented to solve the problems in the standard cosmology (flatness problem, horizon problem, monopole problem, etc). In particular, the chaotic inflation model [9] is the simplest and promising candidate that realizes the inflationary universe. In the chaotic inflation, some scalar field $\phi$, which is called inflaton has a very flat potential, $V(\phi) = \lambda \phi^4/4$ with $\lambda \sim 10^{-13}$. In the chaotic condition of the early universe, the inflaton may have an expectation value much greater than the Planck mass and slowly rolls down to the true minimum of the potential. During the slow rolling epoch, the universe expands exponentially.

When we consider the axion in the chaotic inflationary universe, we confront two serious problems associated with large quantum fluctuations generated during exponential expansion of the universe. One is the domain wall problem [10]. At the epoch of the inflation the axion field $a(x)$ is massless and its fluctuations are given by $\langle a \rangle = H/(2\pi)$ where $H$ is the Hubble constant at that epoch. Since the phase of the Peccei-Quinn scalar $\theta_a$ is related to the $a(x)$ by $\theta_a = a(x)/F_a$, the fluctuations of $\theta_a$ are given by

$$\delta\theta_a = \frac{H}{2\pi F_a}. \qquad (1)$$

In the chaotic inflation with potential $\lambda\phi^4/4$, $H$ is about $10^{14}$ GeV for $\lambda \sim 10^{-13}$ which is required to produce the anisotropies of cosmic microwave background (CMB) observed by COBE–Differential Microwave Radiometer (DMR) [11]. Then, from eq.(1) the fluctuations of the phase $\theta_a$ becomes $O(1)$ for $F_a \lesssim 10^{13}$ GeV, which means that the phase is quite random. Therefore, when the universe cools down to about 1 GeV and the axion potential is formed, axion sits at a different position of the potential at different region of the universe. Since the axion potential has $N$ discrete minima ($N$: colour anomaly), domain walls are produced [12]. The domain wall with $N > 2$ is disastrous because it dominates the density of the universe quickly.

The second problem is that the quantum fluctuations for the axion cause too large anisotropies of CMB [13–15]. Since the axion is massless during the inflation, the axion fluctuations do not contribute to the fluctuations for the total density of the universe. In that sense, the axion fluctuations are isocurvature. After the axion acquires a mass $m_a$, the axion fluctuations becomes density fluctuations given by $\delta\rho_a/\rho_a \sim \delta\theta_a/\theta_a$ which cause the CMB temperature fluctuations $\delta T/T \sim \delta\theta_a/\theta_a$. From eq.(1), the produced CMB anisotropies are $O(1)$ which contradict the observation.

It has been pointed out in ref. [15] that the above two problems are simultaneously solved if the potential of the Peccei-Quinn scalar is very flat. For the flat potential the Peccei-Quinn scalar $\Phi_a$ can take a large expectation value $\sim M_{\rm pl}$ at the epoch of the inflation. Then we should take $\langle \Phi_a \rangle$ as the effective Peccei-Quinn scale instead of $F_a$ and the phase fluctuations are suppressed. Therefore, the production of the domain wall is suppressed and isocurvature fluctuations decreases. However, the isocurvature fluctuations may not be always negligible to the total density fluctuations and hence to the CMB anisotropies. In fact, as is shown in the next section, the isocurvature fluctuations can be comparable with adiabatic ones for a large parameter space. Therefore, it is natural for the axion to have both types of fluctuations in the chaotic inflation scenario.

Since the isocurvature fluctuations give six times larger contribution to the CMB anisotropies at COBE scales [16] than the adiabatic fluctuations, the mixture of isocurvature and adiabatic fluctuations tends to decrease the amplitude of the matter fluctuations if the amplitude is normalize by the COBE–DMR data. This means that relatively high bias parameter is necessary compared with in the pure adiabatic case. In the standard adiabatic cold dark matter scenario, the COBE normalization results in the bias parameter less than 1 which is quite unphysical and also contradicts observations [17–19]. Therefore, the high bias parameter predicted by the model with both adiabatic and isocurvature fluctuations is favoured.

In this paper, we consider the axionic isocurvature fluctuations generated by the chaotic inflation and investigate their cosmological effects on the CMB anisotropies and large-scale structure of the universe.

## II. AXION FLUCTUATIONS

Let us first estimate how large the isocurvature and adiabatic fluctuations are generated in the chaotic inflationary scenario. For a demonstration of our point, we consider a model where the Peccei-Quinn scalar plays a role of an inflaton. The potential for the Peccei-Quinn scalar is given by

$$V(\Phi_a) = \frac{\lambda}{4}(|\Phi_a|^2 - F_a)^2, \qquad (2)$$

with $\lambda \sim 10^{-13}$. Here the axion field $a(x)$ is the phase of $\Phi_a$, namely $\Phi_a = |\Phi_a|e^{ia(x)/F_a}$. We also assume that the axion is dark matter and the density of the axion is equal to the critical density. After the axion acquire a mass, the isocurvature fluctuations with comoving wavenumber $k$ is given by

$$\delta_{\rm iso}(k) \equiv \left(\frac{\delta \rho_a}{\rho_a}(k)\right)_{\rm iso} = \frac{2\delta a}{a} = \frac{\sqrt{2}H}{\Phi_a \theta_a}k^{-3/2}, \qquad (3)$$

where $H$ is the Hubble constant when the comoving wavelength $k^{-1}$ becomes equal to the Hubble radius at the inflation epoch. Since the fluctuations for $\theta_a$ are much smaller than 1 for $\Phi \sim M_{\rm pl}$, the inflation can make $\theta_a$ homogeneous beyond the present horizon of the universe. Therefore, the domain wall problem is solved.

On the other hand the inflaton generates the adiabatic fluctuations which amounts to

$$\delta_{\rm ad}(k) \equiv \left(\frac{\delta \rho}{\rho}(k)\right)_{\rm ad} = \frac{2H^3}{3V'\tilde{H}^2 R(t)^2}k^{1/2}, \qquad (4)$$

where $R(t)$ and $\tilde{H}$ are the scale factor and Hubble constant at arbitrary time $t$. Here we assume the universe is radiation-dominated and the wavelength ($R(t)/k$) is longer the horizon ($\sim \tilde{H}^{-1}$). To compare these two type of fluctuations, it is convenient to take the ratio of the power spectra ($P(k) \equiv \delta(k)^2$) at the horizon crossing, i.e. $k^{-1}R = \tilde{H}^{-1}$, which is written as

$$\alpha \equiv \left.\frac{P_{\rm iso}}{P_{\rm ad}}\right|_{k/R=\tilde{H}} = \frac{9(V')^2}{H^4 \Phi_a^2 \theta_a^2}. \qquad (5)$$

Since the cosmologically interesting scales ($k^{-1} \sim 1$kpc$-$3000Mpc) correspond to the Hubble radius for $\Phi_a \simeq 4M_{\rm pl}$ at the inflation epoch, $\alpha$ is given by

$$\alpha \simeq 2 \times 10^{-3}\theta_a^{-2}. \qquad (6)$$

Furthermore, since the axion is dark matter, the $\theta_a$ is related to the Peccei-Quinn scale $F_a$ by [6]

$$\theta_a \simeq 0.017\left(\frac{\Omega_a h^2}{0.25}\right)^{\frac{1}{2}}\left(\frac{F_a}{10^{15}{\rm GeV}}\right)^{-0.59}, \qquad (7)$$

where $\Omega_a$ is the density parameter of axion at the present epoch and $h$ is the dimensionless Hubble constant normalized by 100km/s/Mpc. Then the ratio $\alpha$ is written as

$$\alpha \simeq 6.24\left(\frac{F_a}{10^{15}{\rm GeV}}\right)^{1.18}\left(\frac{\Omega_a h^2}{0.25}\right)^{-1}. \qquad (8)$$

Therefore, the isocurvature fluctuations are comparable with adiabatic ones for $F_a \gtrsim 10^{14}$ GeV.

It seems natural to take $\theta_a \sim O(1)$ (corresponding to $F_a \sim 10^{12}$ GeV). In this case we have $\alpha \sim 10^{-3}$ and adiabatic fluctuations dominate. However, since $\theta_a$ is homogeneous in the entire universe, we can take $\theta_a$ as a free parameter and we are allowed to take $\theta_a \sim 10^{-2}$ leading to $\alpha \sim O(10)$. This contrasts with the standard cosmology where $\theta_a$ is random in space and the averaged value $\pi/\sqrt{3}$ should be taken. Furthermore, if Peccei-Quinn scalar is independent of the inflaton, the expectation value of $\Phi_a$ at the inflation epoch can be less than the Planck mass depending on the coupling constant of $|\Phi|^4$. In this model we have found that the isocurvature fluctuations comparable with the adiabatic ones, i.e. $\alpha \sim O(1)$, are produced even for $F_a \simeq 10^{12}$GeV (equivalently $\theta_a \sim O(1)$).

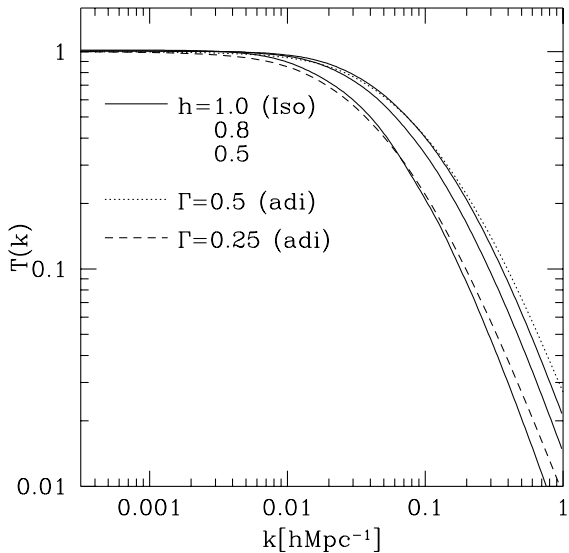

FIG. 1. Matter transfer function $T(k)$ for $\Omega = 1$ pure isocurvature models with $h = 0.5, 0.8$ and $1.0$. Adiabatic CDM models with $\Gamma = 0.5$ and $0.25$ are also plotted.

## III. OBSERVATIONAL CONSTRAINTS

### A. Pure Isocurvature Fluctuations

First we examine models with pure isocurvature fluctuations. Throughout this paper, we only consider models with the total density parameter $\Omega = 1$ and the baryon density parameter $\Omega_b = 0.0125h^{-2}$ from the primordial nucleosynthesis [20]. The matter transfer functions $T(k)$ for different $h$'s are shown in Figure 1 together with adiabatic cold dark matter (CDM) models. The transfer function is defined as $T(k) \equiv \tilde{\delta}(k)/\tilde{\delta}(0)$, where $\tilde{\delta}(k)$ is $k^{3/2}\delta_{\rm iso}(k)$ for isocurvature or $k^{-1/2}\delta_{\rm ad}(k)$ for adiabatic fluctuations. It is well known that the transfer functions of adiabatic CDM models are controlled by a single parameter, $\Gamma \equiv \Omega h$ for low baryon density models [21]. Recent large scale structure observations provide the best fit value as $\Gamma \approx 0.25$ [18,22]. In Figure 1, it should be noticed that the transfer function of the isocurvature model with $h = 0.5$ is very similar to the best fitted adiabatic one. However as we pointed out before, there is a problem of overproducing temperature fluctuations on large scales for isocurvature perturbations. Let us see next the amplitude of mass fluctuations at $8h^{-1}$Mpc, i.e., $\sigma_8$ which is defined as

$$\sigma_8^2 \equiv \frac{1}{2\pi^2} \int \frac{dk}{k} k^3 P(k) \left(\frac{3j_1(kR)}{kR}\right)^2 \bigg|_{R=8h^{-1}{\rm Mpc}}, \quad (9)$$

where $j_1$ is the 1st order spherical Bessel function. The bias parameter $b$ is the inverse of $\sigma_8$: $b = \sigma_8^{-1}$. If we normalize the amplitude of fluctuations to COBE–DMR, the numbers of $\sigma_8$ are 0.11, 0.20 and 0.25 for $h = 0.5$, 0.8 and 1.0, respectively. Here we take the normalization scheme proposed by White and Bunn [23] for the COBE normalization. The observed values of $\sigma_8$ are 0.57 from galaxy clusters survey [17], 0.75 from galaxies and clusters surveys [18] and $0.5 - 1.3$ from peculiar velocity fields [19] if $\Omega = 1$ is assumed. Therefore we can reject pure isocurvature models from these observations.

### B. Adiabatic and Isocurvature Fluctuations

Next we investigate models with admixture of isocurvature and adiabatic perturbations. As we have shown in section II, the amplitude of isocurvature and adiabatic fluctuations are comparable in the chaotic inflationary scenario for a certain range of the initial values of $\theta_a$. Thus it may be interesting to study the cosmological effects of these admixture fluctuations. In linear perturbation theory, isocurvature and adiabatic perturbations are independent solutions. Therefore there is no correlation between these two perturbations. We simply add two power spectra in order to get total one. In Figure 2, we show the values of $\sigma_8$ as a function of $\alpha$ (or $F_a$) for models with $h = 0.5, 0.8$ and $1.0$. As is shown in this figure, we can easily overcome the anti-bias problem of the standard pure adiabatic CDM model by employing admixture models. Assuming the COBE normalization, the value of $\sigma_8$ for the standard CDM, i.e., $\Omega = 1$, $h = 0.5$ and $\Omega_b = 0.05$ with adiabatic perturbations, is 1.4. For admixture models, we can obtain desirable value of $\sigma_8 \simeq 0.5 - 0.8$ for the range of $\alpha \approx 1 - 10$ or $F_a \approx 10^{14-15}$GeV. For the model with $h = 0.5$, $\sigma_8 = 0.57$ for $\alpha = 7.7$ and $\sigma_8 = 0.75$ for $\alpha = 3.8$.

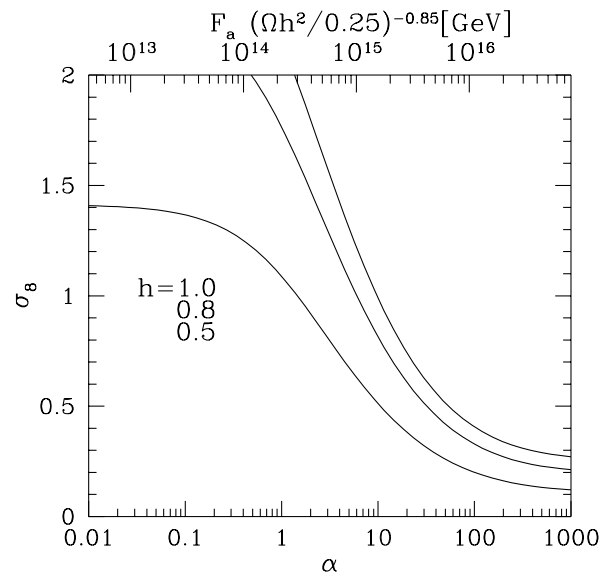

FIG. 2. $\sigma_8$ for models with adiabatic and isocurvature fluctuations as a function of $\alpha$ (lower x-axis) or $F_a \left(\Omega h^2/0.25\right)^{-0.85}$ (upper x-axis). We take $h = 0.5, 0.8$ and $1.0$.

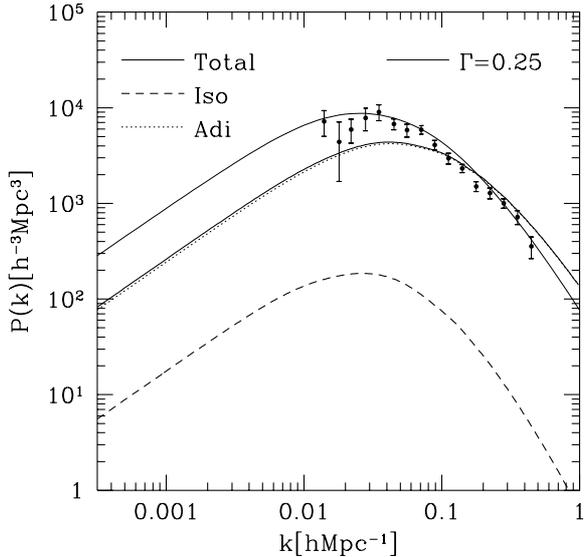

FIG. 3. Matter power spectrum $P(k)$ for the $\Omega = 1, h = 0.5$ and $\Omega_b = 0.05$ model with adiabatic and isocurvature fluctuations. We take $\alpha = 3.75$, i.e., $\sigma_8 = 0.75$. Contributions from isocurvature fluctuations and adiabatic fluctuations are plotted together with total power spectrum. An adiabatic CDM model with $\Gamma = 0.25$ and $\sigma_8 = 0.75$ is also plotted. The observational data are taken from Peacock and Dodds [18].

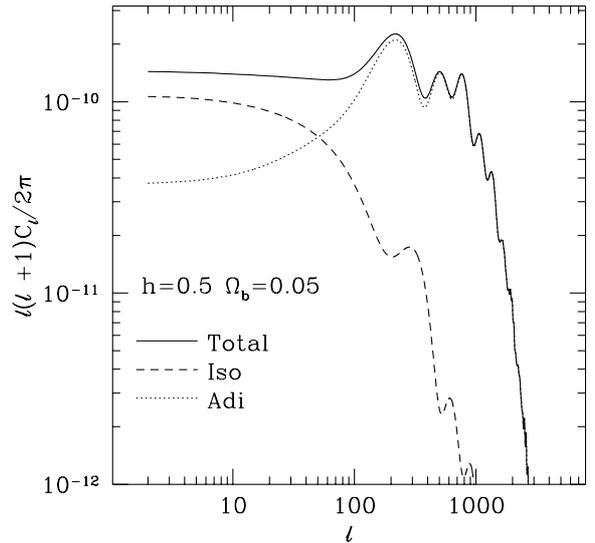

FIG. 4. Power spectrum of CMB anisotropies $C_\ell$ for the $\Omega = 1, h = 0.5$ and $\Omega_b = 0.05$ model with adiabatic and isocurvature fluctuations as a function of multipole components $\ell$. We take $\alpha = 3.75$, i.e., $\sigma_8 = 0.75$. Contributions from isocurvature fluctuations and adiabatic fluctuations are plotted together with total power spectrum.

The matter power spectrum of the model with $h = 0.5$ and $\alpha = 3.8$ i.e., $\sigma_8 = 0.75$ is shown in Figure 3. If $\sigma_8 \gtrsim 0.5$, the difference between pure adiabatic and admixture of adiabatic and isocurvature fluctuations is very small. Therefore, the same problem with pure adiabatic CDM models arises. Namely, it is difficult to obtain the shape fitted with observations if we employ $h \gtrsim 0.5$. However, a recent analysis of velocity fields [24], which have much sensitivity on larger scales, suggests that the turnover point of the power spectrum is smaller than we thought before. Their best fitted shape is $\Gamma \simeq 0.5$. Therefore, it is premature to rule out the model merely from the shape of the power spectrum.

In Figure 4, CMB anisotropy multipole moments $C_\ell = <|a_{\ell m}|^2>$ are shown. Here $\delta T/T = \sum_{\ell m} a_{\ell m} Y_{\ell m}$ with $Y_{\ell m}$ being spherical harmonics. There is a clear distinction between pure adiabatic and admixture spectra. It might be possible to determine $\alpha$ from future experiments by a new satellite or a balloon with long duration flight.

## IV. CONCLUSION AND REMARKS

We study cosmological implication of axion in the chaotic inflation scenario. By assuming very flat potential for the Peccei-Quinn scaler, we can solve overproduction problems of domain walls and of CMB anisotropies. The simplest model where the Peccei-Quinn scaler plays a role of an inflaton of the chaotic inflation is investigated. This scaler filed produces both adiabatic and isocurvature fluctuations. From recent observations of large scale structure and CMB anisotropies, models with pure isocurvature fluctuations (or negligible amount of adiabatic fluctuations) are ruled out. The preferable value of $F_a$ for the desired bias parameter ($b \sim 2$) is about $10^{15}$GeV which happens to be a GUT scale.

**Note added:** After finishing up this paper, we have found a paper by Stomper et al. [25] which has also discussed the cosmological consequences of the admixture fluctuations. However explicit models for producing both adiabatic and isocurvature fluctuations have not been considered in their paper.

## ACKNOWLEDGMENTS

We would like to thank M. Sasaki for useful comments.